\documentclass[aps,prd,twocolumn,groupedaddress,showpacs]{revtex4}

\usepackage{graphicx}
\usepackage{dcolumn}
\usepackage{bm}
\usepackage{amsmath}
\usepackage{epstopdf}
\usepackage{amsfonts}
\usepackage{amssymb}
\usepackage{hyperref}
\usepackage{xcolor}

\usepackage[utf8]{inputenc}
\usepackage{natbib}

\providecommand{\U}[1]{\protect\rule{.1in}{.1in}}

\newcommand{\baa}{\begin{align}}
\newcommand{\eaa}{\end{align}}

\newcommand{\be}{\begin{equation}}
\newcommand{\ee}{\end{equation}}
\newcommand{\bea}{\begin{eqnarray}}
\newcommand{\eea}{\end{eqnarray}}

\begin{document}

\title{Quasinormal spectra of scale-dependent Schwarzschild-de Sitter black holes}


\author{Grigoris Panotopoulos}

\affiliation{
Centro de Astrof\'{\i}sica e Gravita{\c c}{\~a}o, Departamento de F{\'i}sica, Instituto Superior T\'ecnico-IST, Universidade de Lisboa-UL, 
Av. Rovisco Pais, 1049-001 Lisboa, Portugal.
}

\email{grigorios.panotopoulos@tecnico.ulisboa.pt}

\author{\'Angel Rinc\'on}

\affiliation{
Instituto de F\'isica, Pontificia Universidad Cat\'olica de Valpara\'iso,
\mbox{Avenida Brasil 2950, Casilla 4059, Valpara\'iso, Chile.}
}

\email{angel.rincon@pucv.cl}

\date{\today}


\begin{abstract}
We compute the quasinormal spectra for scalar, Dirac and electromagnetic perturbations of the Schwarzschild-de Sitter geometry in the framework of scale-dependent gravity, which is one of the current approaches to quantum gravity. Adopting the widely used WKB semi-classical approximation, we investigate the impact on the spectrum of the angular degree, the overtone number as well as the scale-dependent parameter for fixed black hole mass and cosmological constant. We summarize our numerical results in tables, and for better visualization, we show them graphically as well. All modes are found to be stable. Our findings show that both the real part and the absolute value of the imaginary part of the frequencies increase with the parameter $\epsilon$ that measures the deviation from the classical geometry. Therefore, in the framework of scale-dependent gravity the modes oscillate and decay faster in comparison with their classical counterparts.
\end{abstract}



\maketitle

\section{Introduction}\label{Intro}

Many astonishing predictions of Einstein's general relativity \cite{GR} (GR) have been confirmed over the last 100 years or so, since its formulation until today. On the one hand, the classical tests and solar system tests \cite{tests1,tests2,tests3}, and on the other hand, more recently, the first image of a black hole \cite{L1,L2,L3,L4,L5,L6} combined with the direct detections of gravitational waves \cite{ligo1,ligo2,ligo3,ligo4,ligo5}, from mergers of black hole (BH) and neutron star binary systems, have shown in the most convincing way that GR is a remarkable theory of gravity, and also they have provided us with the strongest evidence so far that black holes do exist in Nature and that they merge. Astrophysical BHs are supposed to be formed during the final stages of massive stars, primordial BHs may be formed in the early Universe from density inhomogeneities, and mini-black holes may be formed at colliders or in the atmosphere of the earth in TeV-scale gravity scenarios in D-brane constructions of the Standard Model \cite{miniBH1,miniBH2,miniBH3}.

\smallskip

As successful as it may be as a classical theory, at quantum level technically speaking GR falls into the class of non-renormalizable theories. Although by now we know how to extract quantum predictions from a non-renormalizable theory using the techniques of effective field theory (to which GR fits perfectly) \cite{Donoghue:1994dn}, the problem remains. As of today, the quest for a consistent quantum theory of gravity is still an open task in modern theoretical physics. In the context of theories beyond classical GR, one may mention at least two approaches, namely either scale-dependent gravity \cite{Contreras:2013hua,Koch:2015nva,SD1,Rincon:2017ayr,Hernandez-Arboleda:2018qdo,Contreras:2018gct,Canales:2018tbn,Rincon:2019zxk,Rincon:2018lyd,Contreras:2019fwu,
Fathi:2019jid,PRL2,Contreras:2018swc,Sendra:2018vux,SD2,SD3,SD4,SD5,SD6,SD7,ourprd} or improvement asymptotically safe gravity \cite{Bonanno:2000ep,Bonanno:2001xi,Reuter:2003ca,Koch:2014cqa,Gonzalez:2015upa}. In both approaches the basic quantities that enter into the action defining the model at hand, such as Newton's constant, cosmological constant etc, become scale dependent quantities. This does not come as a surprise of course, since scale dependence at the level of the effective action is a generic feature of ordinary quantum field theory. 

\smallskip

Realistic black holes in Nature are in a constant interaction with their environment. When a black hole is perturbed, it responds by emitting gravitational waves. Quasinormal modes (QNMs) are characteristic frequencies with a non-vanishing imaginary part, which encode the information on how black holes relax after the perturbation has been applied. The QN frequencies depend on the details of the geometry and on the type of the perturbation (scalar, vector, tensor or fermionic), but not on the initial conditions. Black hole perturbation theory \cite{wheeler,zerilli1,zerilli2,zerilli3,moncrief,teukolsky} and QNMs become relevant during the ring down phase of a black hole merger, the stage where a single distorted object is formed, and where the geometry of space-time undergoes dumped oscillations due to the emission of gravitational waves. Thanks to gravitational wave Astronomy we have now a powerful tool at our disposal to test gravitation under extreme conditions. For excellent reviews on the topic see \cite{review1,review2,review3}, and also the Chandrasekhar's monograph \cite{chandra}, which is the standard textbook on the mathematical aspects of black holes. 

\smallskip 

The current cosmic acceleration \cite{Riess:1998cb,Perlmutter:1998np} motivates the study of space-times with a positive cosmological constant. The maximally symmetric space-time of Einstein's field equations with a positive cosmological constant without the presence of matter fields is the Schwarzschild-de Sitter geometry, characterized by the cosmological constant itself and the mass of the black hole. The quasinormal frequencies of the Schwarzschild-de Sitter in four dimensions and higher dimensions have been studied e.g. in \cite{lemos,molina,Zhidenko:2003wq,Konoplya:2004uk,LopezOrtega:2006my,Kodama:2003jz,panotop1} and references therein. Given the importance and relevance of the QNMs of black holes in gravitational wave Astronomy, it would be interesting to see what kind of spectra are expected from scale-dependent black holes.

\smallskip

In black hole physics the impact of the SD scenario on properties of black holes has been studied over the last years, and it has been found that the scale dependence modifies the horizon, the thermodynamics as well as the quasinormal spectra of classical black hole backgrounds \cite{SD1,SD2,SD3,SD4,SD5,SD6,SD7,ourprd}. However, to the best of our knowledge, the QN modes of the four-dimensional scale-dependent Schwarzschild-de Sitter geometry have not been studied yet. Therefore, in the present work we propose to compute the QN spectra for scalar, Dirac, and electromagnetic perturbations of the SD Schwarzschild-de Sitter in four dimensions, filling thus a gap in the literature, and also making a direct comparison with the frequencies of the classical Schwarzschild-de Sitter solution.

\smallskip

The plan of our work is the following: In the next section we present the wave equations with the corresponding effective potential barrier for scalar, Dirac and electromagnetic perturbations, while in section three we briefly review the effective field equations in the framework of scale-dependent gravity. In section four we compute the quasinormal frequencies adopting the WKB approximation of 6th order, and we discuss our results. Finally, we summarize our work with some concluding remarks in section five. We adopt the mostly positive metric signature $(-,+,+,+)$, and we work in geometrical units where the universal constants are set to unity, $c=1=G_0$.

\section{Wave equations for massless scalar, Dirac and electromagnetic perturbations} \label{Wave}

In this section we present the wave equation and the corresponding effective potential barrier for massless scalar, Dirac and electromagnetic perturbations. We consider a fixed gravitational background (static, spherically symmetric) in Schwarzschild-like coordinates $t,r,\theta,\phi$ of the form
\begin{equation} \label{background}
ds^2 = -f(r) dt^2 + f(r)^{-1} dr^2 + r^2 (d \theta^2 + \sin^2(\theta) d \phi^2)
\end{equation}
with a given lapse function $f(r)$. Then we perturb that geometry by throwing in a probe field, and we study its propagation.

\subsection{Spin zero case (scalar field)}

The wave equation of a massless minimally coupled scalar field $\Phi$ is given by the usual Klein-Gordon equation, namely
\cite{coupling,p11,p12}
\begin{equation}
\frac{1}{\sqrt{-g}} \partial_\mu 
\Bigl(
\sqrt{-g} g^{\mu \nu} \partial_\nu
\Bigl) 
\Phi = 0
\end{equation}
In order to solve the previous equation, we apply the separation of variables making the usual ansatz:
\begin{equation}\label{separable}
\Phi(t,r,\theta,\phi) = e^{-i \omega t} \frac{\psi(r)}{r} Y_l^m (\theta, \phi)
\end{equation}
where $Y_l^m$ are the usual spherical harmonics, and $\omega$ is the frequency to be determined. 
Making the previous ansatz it is straightforward to obtain for the radial part a Schr{\"o}dinger-like equation
\begin{equation}
\frac{\mathrm{d}^2 \psi}{\mathrm{d}x^2} + [ \omega^2 - V(x) ] \psi = 0
\end{equation}
with $x$ being the tortoise coordinate, i.e.,
\begin{equation}
x  \equiv  \int \frac{\mathrm{d}r}{f(r)}
\end{equation}
Finally, the effective potential barrier is given by
\begin{equation}
V_s(r) = f(r) \: \left(\frac{l (l+1)}{r^2}+\frac{f'(r)}{r} \right), \; \; \; \; l \geq 0
\end{equation}
where the prime denotes differentiation with respect to $r$, and $l$ is the angular degree.

\smallskip

\subsection{Spin one-half case (Dirac field)}

Fermions are coupled to gravity in the vierbein formalism. Let us here briefly review the formalism, and present the wave equation with the corresponding potential following \cite{Destounis:2018qnb}. First we introduce the tetrad or vierbein, $e_a^\mu$, defined by
\begin{equation}
e_\mu^a e_\nu^b \eta_{ab} = g_{\mu \nu}
\end{equation}
with $\eta_{ab}$ being the flat Minkowski metric tensor. The vierbein carries two kinds of indices, namely a flat index $a$ as well a space-time index $\mu$, and it may be viewed as the "square root" of the metric tensor $g_{\mu \nu}$. Next we define curved Dirac matrices as follows
\begin{equation}
G^\mu \equiv e_a^\mu \gamma^a
\end{equation}
with the following property
\begin{eqnarray}
\{ \gamma^a, \gamma^b \} & = & -2 \eta^{ab} \\
\{ G^\mu, G^\nu \} & = & -2 g^{ab}
\end{eqnarray}
where $\gamma^a$ are the usual Dirac matrices from the usual relativistic quantum mechanics in flat space-time. Finally, introducing the spin connection $\omega_{ab \mu}, \Gamma_\mu$
\begin{eqnarray}
\Gamma_\mu & = & -\frac{1}{8} \omega_{ab\mu} [\gamma^a, \gamma^b] \\
\omega_{ab \mu} & = & \eta_{ac} [e_\nu^c e_b^\lambda \Gamma^\nu_{\mu \lambda} - e_b^\lambda \partial_\mu e_\lambda^c]
\end{eqnarray}
where $\Gamma^\nu_{\mu \lambda}$ are the Christoffel symbols, the Dirac equation for a spin one-half fermion $\Psi$ in curved space-time is given by
\begin{equation}
(i G^\mu D_\mu - m_f) \Psi = 0
\end{equation}
where $m_f$ is the mass of the fermion, while the covariant derivative is defined to be $D_\mu \equiv \partial_\mu + \Gamma_\mu$.

\smallskip

For Dirac fermions we separate variables as before, where now we use the spinor spherical harmonics \cite{Finster:1998ak}, and two radial parts
\begin{align}
r^{-1} \: f(r)^{-1/4} \: F(r) 
\\
r^{-1} \: f(r)^{-1/4} \: i G(r)
\end{align}
for the upper and lower components of the Dirac spinor $\Psi$, respectively. In 
the massless limit, $m_f=0$, one obtains the following equations \cite{Destounis:2018qnb}
\begin{eqnarray}
\frac{dF}{dx} - W F + \omega G & = & 0 \\
\frac{dG}{dx} + W G - \omega F & = & 0
\end{eqnarray}
where $x$ is the tortoise coordinate as before, while the function $W$ is given by \cite{Destounis:2018qnb}
\begin{equation}
W = \frac{\xi \sqrt{f}}{r}
\end{equation}
where $\xi = \pm (j+1/2)=\pm 1, \pm 2,...$, with $j$ being the total angular momentum, $j=l \pm 1/2$ \cite{Destounis:2018qnb}. Finally, the two coupled equations for $F,G$ can be combined to obtain a wave equation for each one of the following form
\begin{eqnarray}
\frac{d^2F}{dx^2} + [\omega^2 - V_+] F & = & 0 \\
\frac{d^2G}{dx^2} + [\omega^2 - V_-] G & = & 0
\end{eqnarray}
where the potentials are found to be
\begin{equation}
V_{\pm} = W^2 \pm \frac{dW}{dx}
\end{equation}
In the language of supersymmetry, the potentials $V_-,V_+$ are superpartners, as they are derived from a superpotential, and consequently they yield the same spectra, see e.g. \cite{Cooper:1994eh}. Therefore, in the following we shall work with the plus sign and the wave equation for $F(r)$.

\smallskip

\subsection{Spin one case (Maxwell field)}

Electromagnetic perturbations are governed by Maxwell's equations \cite{Cardoso:2001bb}
\begin{equation}
F^{\mu \nu}_{;\nu} = 0, \; \; \; \; \; \; F_{\mu \nu} \equiv \partial_\mu A_\nu - \partial_\nu A_\mu
\end{equation}
where $A_\mu$ is the Maxwell potential, $F_{\mu \nu}$ is the electromagnetic field strength, and a semi-colon denotes covariant derivative. Since the background geometry is spherically symmetric, one can expand the Maxwell potential in vector spherical harmonics, see for instance \cite{Cardoso:2001bb}.
Following a similar separation of variables as before, one obtains a Schr{\"o}dinger-like equation with the following effective potential barrier for electromagnetic perturbations \cite{lemos,Cardoso:2001bb}
\begin{equation}
V_{EM}(r) = f(r) \: \left( \frac{l (l+1)}{r^2} \right), \; \; \; \; l \geq 1
\end{equation}
Therefore, to study the QN spectra of a given geometry with lapse function $f(r)$, one needs to solve the Regge-Wheeler equation
\begin{equation}
\frac{\mathrm{d}^2 \psi}{\mathrm{d}x^2} + [ \omega^2 - V(x) ] \psi = 0
\end{equation}
where the effective potential is given by the following expressions depending on the type of perturbation
\begin{align}
V_{s=0} & =  f(r) \: \left(\frac{l (l+1)}{r^2}+\frac{f'(r)}{r} \right)
 \\
V_{s=1} & =  f(r) \: \left( \frac{l (l+1)}{r^2} \right) 
\\
V_{s=1/2} &= W^2 \pm \frac{dW}{dx}, \; \; \; \; \; W = \frac{\xi \sqrt{f}}{r}
\end{align}

\section{Scale-dependent Schwarzschild-de Sitter geometry}

\subsection{Classical geometry}

In order to appreciate the corrections induced in the scale-dependent scenario, we first review very briefly the classical solution. We then start by considering Einstein's field equations with a positive cosmological constant, $\Lambda_0 > 0$, in a four-dimensional space-time
\begin{equation}
R_{\mu \nu} - \frac{1}{2} R g_{\mu \nu} + \Lambda_0 g_{\mu \nu} = 0
\end{equation}
with $g_{\mu \nu}$ being the metric tensor, $R_{\mu \nu}$ the Ricci tensor and $R$ the Ricci scalar.

The gravitational field produced by a point-like mass $M$ is given by the well-known Schwarzschild-de Sitter solution \cite{Tangherlini}
\begin{equation}
ds^2 = - f_{0}(r) dt^2 + f_{0}(r)^{-1} dr^2 + r^2 d\Omega^2
\end{equation}
where $r$ is the radial coordinate, $d\Omega^2$ is the usual line element of the unit 2-sphere,
\begin{align}
d\Omega^2 = d \theta^2 + \sin^2 \theta d \phi^2
\end{align}
while the classical lapse function is given by
\begin{equation}
f_{0}(r) = 1-\frac{2 M}{r} - \frac{1}{3}\Lambda_0 r^2
\end{equation}
where the sub index $0$ denotes classical quantities. To compute the horizon one needs to solve the algebraic equation $f(z)=0$, which leads to the cubic equation
\begin{equation}
z^3 - \frac{3 z}{\Lambda_0} + \frac{6M}{\Lambda_0} = 0
\end{equation}
According to the general theory of the cubic equation, when the descriminant $D$ computed by
\begin{eqnarray} 
D & \equiv & R^2 + Q^3 \\
R & = & - \frac{3M}{\Lambda_0} \\
Q & = & - \frac{1}{\Lambda_0}
\end{eqnarray}
is negative, all roots are real and unequal. In the case of the classical Schwarzschild-de Sitter geometry, the condition $D < 0$ is equivalent to
\begin{equation}
\Lambda_0 M^2 < \frac{1}{9}
\end{equation}
and there are i) a real negative root, and ii) two real positive roots corresponding to an event horizon $r_H$, and a cosmological horizon $r_c > r_H$. In the special cases $\Lambda_0=0$ and $M=0$, the Schwarzschild solution and the 
maximally symmetric space-time corresponding to de-Sitter solution, respectively, are recovered.

\smallskip

\subsection{Scale-dependent geometry}

In what follows, we will summarize the main features of the scale-dependent scenario. The different approaches to deal with theories where quantum and classical elements can coexist are quite diverse. However, most of them share one feature in particular, namely they show a non-trivial scale dependence, as is also reported in most other quantum field theories. Thus, the starting point in the scale-dependent formalism is the well-known average effective action, which accounts for both classical and quantum features. 

\smallskip

Avoiding unnecessary details (see \cite{Contreras:2013hua,Koch:2015nva,SD1,Rincon:2017ayr,Hernandez-Arboleda:2018qdo,Contreras:2018gct,Canales:2018tbn,Rincon:2019zxk,Rincon:2018lyd,Contreras:2019fwu,
Fathi:2019jid,PRL2,Contreras:2018swc,Sendra:2018vux,SD2,SD3,SD4,SD5,SD6,SD7,ourprd}), we can take advantage of the scale-dependent version of the Einstein-Hilbert action with a non-vanishing cosmological constant. In this approach, the fundamental object, $\Gamma[g_{\mu \nu}, k]$, replaces the classical action $I[g_{\mu \nu}]$, and the most notorious difference emerges due the inclusion of scale-dependent couplings. Symbolically we can write
\begin{align}\label{coupling}
\{A_0, B_0, (\cdots)_0, Z_0\} \ \rightarrow \ \{A_k, B_k, (\cdots)_k, Z_k\}
\end{align}
where the left-hand side of Eq~\eqref{coupling} represents the classical coupling, and the right-hand side collects the scale-dependent couplings. Also, it is essential to point out that $k$ is an arbitrary renormalization scale connected to some physical variables later on. Besides, as the solutions obtained in the scale-dependent scenario also contain GR as a limiting case, any modification should guarantee that the scale-dependent solution, obtained from the effective action, boils down to the classical solution when the appropriate limit is taken.
In the SD scenario \cite{Rincon:2017ayr} we will consider three different contributions accounted into the following action:
\begin{align} \label{action}
S[g_{\mu \nu},k] \equiv S_{\text{EH}} + S_{\Lambda} + S_{\text{SD}}
\end{align} 
where the first term, $S_{\text{EH}}$, is given by the usual Einstein-Hilbert action, the second term, $S_{\Lambda}$, is associated with the cosmological constant, and the third term, $S_{\text{SD}}$, encodes the scale-dependent sector.
Usually, this kind of system can be solved if we know the concrete form of the coupling constants (see also \cite{SD1} for details), or, for example, from a background independent integration of the functional renormalization group \cite{Wetterich:1992yh,Dou:1997fg,Souma:1999at,Reuter:2001ag}. Here we shall follow an alternative route: we know that the functional form of $k(\cdots)$ can be written as a function of the radial coordinate (in spherical symmetry). Thus, we will take advantage of this fact to promote both $\Lambda_0$ and Newton's constant $G_0$ to scale-dependent quantities
\begin{eqnarray} 
G_0 & \rightarrow & G(r) \\
\Lambda_0 & \rightarrow & \Lambda(r)
\end{eqnarray}
the effective field equations, obtained from a variation of \eqref{action} with respect to $g_{\mu \nu}(x)$, are the following \cite{Rincon:2017ayr}:
\begin{align}
R_{\mu \nu} - \frac{1}{2}R g_{\mu \nu} + \Lambda(r) g_{\mu \nu} = - \Delta t_{\mu \nu}
\end{align}
where the G-varying part $\Delta t_{\mu \nu}$ is computed to be \cite{Rincon:2017ayr}
\begin{equation}
\Delta t_{\mu \nu} = G_k \Bigl(g_{\mu \nu} \Box - \nabla_\mu \nabla_\nu\Bigl) G_k^{-1}
\end{equation}
What is more, there is an additional differential equation for $G(r)$, which is the 
following \cite{Rincon:2017ayr}
\begin{equation}
2 \frac{G(r)''}{G(r)'} - 4 \frac{G(r)'}{G(r)} = \Bigl( \mathrm{ln}\left( \frac{f(r)}{g(r)} \right) \Bigl)'
\end{equation}
where $f(r),g(r)$ are the two unknown components of the metric tensor in SD gravity:
\begin{equation}
ds^2 = - f(r) dt^2 + g(r)^{-1} dr^2 + r^2 (d \theta^2 + \sin^2(\theta) d \phi^2)
\end{equation}
If, however, for simplicity one assumes the usual Schwarzschild ansatz, $f(r) = g(r)$,
the equation for $G(r)$ may be integrated immediately to obtain the known solution already obtained in previous works \cite{Rincon:2018lyd,SD7}
\begin{equation}
G(r) = \frac{G_0}{1 + \epsilon r}
\end{equation}
where $\epsilon$ is the SD parameter. Clearly, the classical result is recovered in the limit
$\epsilon \rightarrow 0$. 

\smallskip

\begin{figure*}[ht]
\centering
\includegraphics[width=0.48\textwidth]{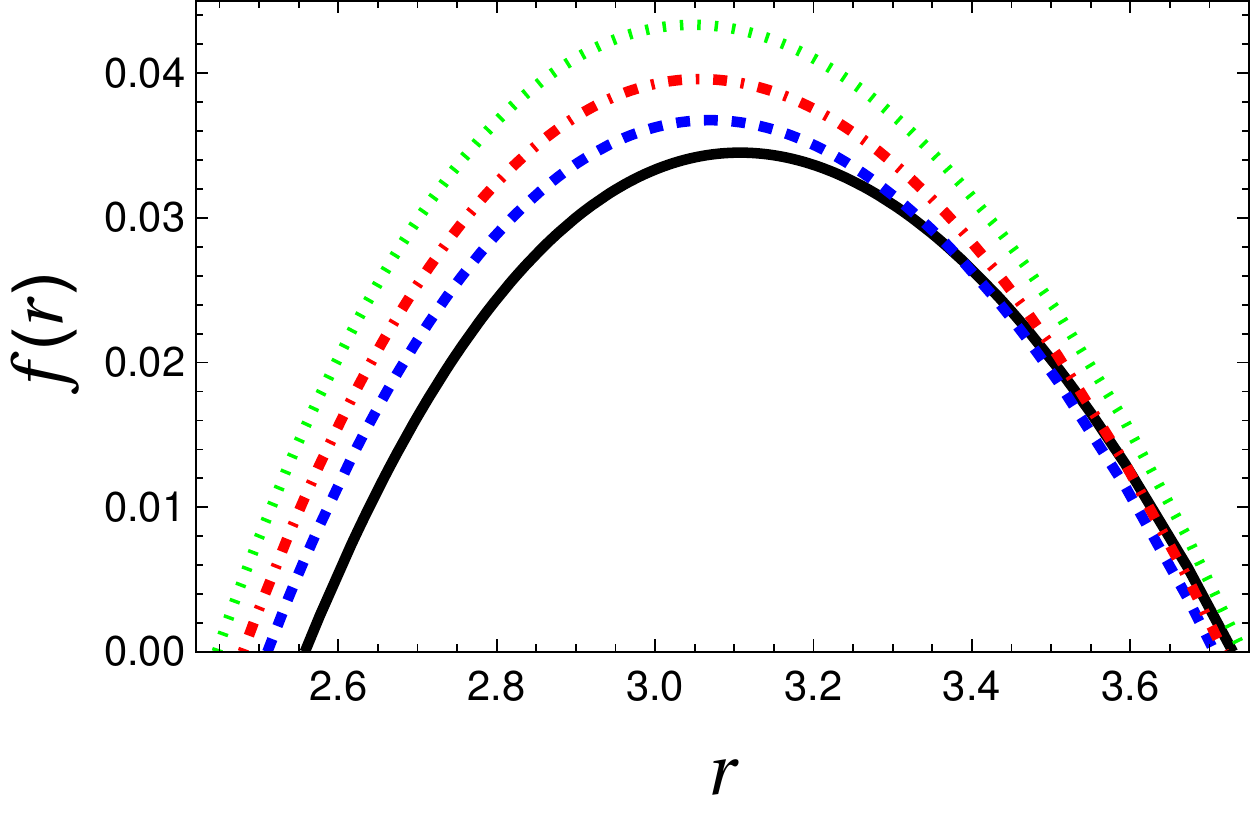}   
\ \ \
\includegraphics[width=0.48\textwidth]{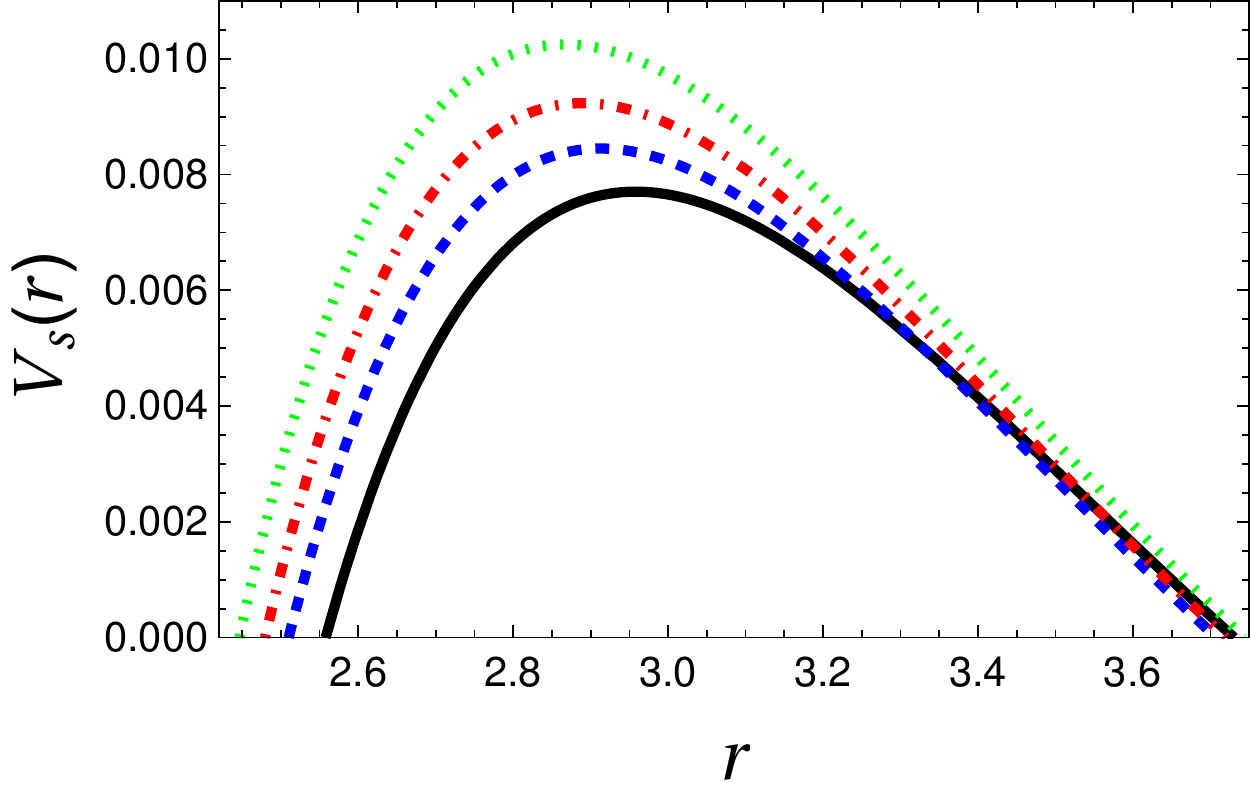}   
\\
\caption{
{\bf{LEFT:}} Lapse function $f(r)$ vs radial coordinate $r$ for $M=1$ and $\Lambda_0=0.1$ in four different cases for the scale-dependent parameter $\epsilon$: 
i)   $\epsilon = 0$ (classical solution, solid black line), 
ii)  $\epsilon = 0.010$ (dashed blue line), 
iii) $\epsilon = 0.015$ (dot-dashed red line), and finally 
iv)  $\epsilon = 0.020$ (dotted green line).
{\bf{RIGHT:}} Effective potential barrier $V_s(r)$ for the scalar case, taking $M=1$, $\Lambda_0=0.1$ and $l=1$ in four different cases of the scale-dependent parameter $\epsilon$ as follows:  
i)   $\epsilon = 0$ (solid black line), 
ii)  $\epsilon = 0.010$ (dashed blue line), 
iii) $\epsilon = 0.015$ (dot-dashed red line), and finally 
iv)  $\epsilon = 0.020$ (dotted green line).
}
\label{fig:Lapse-Potential}
\end{figure*}

With Newton's constant already determined, using the effective field equations it is not difficult to verify that the r-varying cosmological constant $\Lambda(r)$ as well as the lapse function $f(r)$ are found to be
\begin{align} 
\Lambda(r) =  
\ \Lambda_0 & + \frac{\epsilon r}{1 + \epsilon r }\Lambda_0 + 
\epsilon^2
\Bigg[
\frac{3 G_0 M}{r}
\frac{(1 + 12 \epsilon r (1 + \epsilon r ) )}{(1 + \epsilon r)^2}
\nonumber
\\
&
+
\frac{2-3 r \epsilon  (6 r \epsilon +5)}{2 (r \epsilon +1)^2}
-
3 (1 + 6 G_0 M \epsilon) \times
\nonumber
\\
&
\times
\Bigg(
\frac{1 + 2 \epsilon r}{1 + \epsilon r }
\Bigg) 
\ln \left(1 + \frac{1}{r \epsilon } \right) 
\Bigg]
\end{align}
\begin{align}
\begin{split}
f(r)  =  f_0(r) + & \frac{1}{2}\epsilon 
\Bigg[
6 G_0 M - 2r + 3 r \epsilon  (r - 4 G_0 M)
\\
& +
 2 r^2 \epsilon  (1 + 6 G_0 M \epsilon) \ln \left( 1 + \frac{1}{r \epsilon } \right)
\Bigg]
\end{split}
\end{align}
and clearly
\begin{align} 
\lim_{\epsilon \rightarrow 0} \Lambda(r) &\rightarrow \Lambda_0
\\
\lim_{\epsilon \rightarrow 0} f(r) &\rightarrow f_0(r)
\\
\lim_{\epsilon \rightarrow 0} G(r) &\rightarrow G_0 \equiv 1
\end{align}
Finally, for $\epsilon$ sufficiently small, one may use a Taylor expansion of second order in powers of it, namely
\begin{align} 
\begin{split}
f(r) \approx & \ f_{0}(r) - \left(1 - \frac{3M}{r}\right) \: (\epsilon r)  + 
\\
&
\left(\frac{3}{2} -\frac{6M}{r} - \mathrm{ln}(\epsilon r) \right) \: (\epsilon r)^2 + \mathcal{O}(\epsilon^3)
\end{split}
\end{align}
In Fig.~\ref{fig:Lapse-Potential} we show the lapse function $f(r)$ (left panel) and the scalar effective potential barrier (right panel) for the classical solution as well as for three different values of $\epsilon$, namely $\epsilon=0.01, 0.015, 0.02$.

Unfortunately the form of the lapse function makes it impossible to obtain analytical expressions for the roots of the algebraic equation $f(r)=0$. For sufficiently small $\epsilon$, however, the above expansion may be used to obtain the condition for the existence of three real and unequal roots.
Thus, up to first order in $\epsilon$ we find
\begin{align}
\Lambda_0 M^2 < \frac{1}{9} + \mathcal{O}(\epsilon^2)
\end{align}
which means that the classical bound is still valid.


\section{QNM spectra of SD Schwarzschild-de Sitter BH in the WKB approximation}

Exact analytical calculations are always desirable, since it is only in this case where 
the complete parameter space can be fully explored. Regarding QN spectra of black holes,
exact analytic expressions exist in very few cases, e.g. either when the effective potential barrier takes the form of the P{\"o}schl-Teller potential \cite{potential,ferrari,lemos,molina,panotop1}, or when the radial part of the wave function may be recast into the Gauss' hypergeometric function \cite{exact1,exact2,exact3,exact4,exact5,exact6}. More generically, over the years several different methods to compute the QNMs of black holes have been developed, such as the Frobenius method, generalization of the Frobenius series, fit and interpolation approach, method of continued fraction etc. For more details the interested reader may want to consult e.g. \cite{review3}. In particular, semi-analytical methods based on the Wentzel-Kramers-Brillouin (WKB) approximation \cite{wkb1,wkb2,wkb3} are among the most popular ones, and they have been extensively applied to several cases. For higher order WKB corrections, and recipes for simple, quick, efficient and accurate computations see \cite{Opala,Konoplya:2019hlu,RefExtra2}.
For a partial list see for instance \cite{paper1,paper2,paper3,paper4,paper5,paper6}, and for more recent works \cite{paper7,paper8,paper9,paper10,Rincon:2018sgd,Rincon:2020iwy}, and references therein.

Within the WKB approximation the QN frequencies are computed by
\begin{equation}
\omega_n^2 = V_0+(-2V_0'')^{1/2} \Lambda(n) - i \nu (-2V_0'')^{1/2} [1+\Omega(n)]
\end{equation}
where $n=0,1,2...$ is the overtone number, $\nu=n+1/2$, $V_0$ is the maximum of the effective potential, $V_0''$ is the second derivative of the effective potential evaluated at the maximum, while $\Lambda(n), \Omega(n)$ are complicated expressions of $\nu$ and higher derivatives of the potential evaluated at the maximum, and can be seen e.g. in \cite{paper2,paper7}. 
Here we have used the Wolfram Mathematica \cite{wolfram} code with WKB at any order from one to six, which has been presented in \cite{code},  and it can be downloaded from https://g00.gl/hykYGL. 
It is known that within the WKB approach the best results are obtained for higher angular degrees and low overtone number, see e.g. the numerical values shown in Tables II, III, IV and V of \cite{Opala}. Therefore in this work we shall consider $l=6,7,8,9,10$ and $n=0,1,2$ for the scalar and electromagnetic perturbations, while for Dirac perturbations we consider the fundamental mode of $\xi=7,8,9,10,11$.


\begin{table*} 
\centering
\caption{QN frequencies for scalar perturbations, setting $M=1$ and $\Lambda_0=0.1$ for three different values of the scale-dependent parameter $\epsilon$. For comparison reasons, we also show the frequencies of the classical geometry, i.e., when $\epsilon =0$.
The modes are presented as follows: 
i) classical modes (without parenthesis), 
ii) scale-dependent modes with $\epsilon = 0.01$ (in parenthesis),
iii) scale-dependent modes with $\epsilon=0.015$ (in brackets), and finally
iv) scale-dependent modes with $\epsilon=0.02$ (in curly brackets). 
}
\resizebox{2\columnwidth}{!}{%
\begin{tabular}{cccccc}
\hline
$n$ &  $l=6$ & $l=7$ & $l=8$ & $l=9$ & $l=10$\\
\hline
     &  0.39349\, -0.0304656 i & 0.454634\, -0.030456 i & 0.515703\, -0.0304501 i & 0.576729\, -0.0304459 i & 0.637722\, -0.0304429 i  \\
0    &  (0.411628\, -0.0318694 i) & (0.475575\, -0.0318598 i) & (0.539456\, -0.0318531 i) & (0.603291\, -0.0318484 i) & (0.667091\, -0.0318451 i) \\
     & [0.429689\, -0.0332691 i] & [0.49644\, -0.033258 i] & [0.563117\, -0.0332507 i] & [0.629748\, -0.0332455 i] & [0.696343\, -0.0332418 i]\\
     & \{0.451879\, -0.0349884 i \} & \{0.522067\, -0.0349759 i \} & \{ 0.592181\, -0.0349676 i \} & \{ 0.662247\, -0.0349617 i \} & \{0.732275\, -0.0349575 i\} \\
\hline
     &  0.39318\, -0.0914102 i & 0.45444\, -0.0913634 i & 0.515507\, -0.091351 i & 0.576549\, -0.0913392 i & 0.637555\, -0.0913304 i \\
1    &  (0.411359\, -0.0956035 i) & (0.475308\, -0.0955831 i) & (0.539219\, -0.0955625 i) & (0.60309\, -0.0955462 i) & (0.666899\, -0.0955374 i) \\ 
     &  [0.429343\, -0.0998119 i] & [0.496158\, -0.099774 i] & [0.562849\, -0.0997559 i] & [0.629518\, -0.0997381 i] & [0.69613\, -0.0997276 i] \\
     &  \{0.451479\, -0.104971 i\} & \{ 0.521721\, -0.104932 i\} & \{0.59187\, -0.104907 i\} & \{0.661979\, -0.104887 i\} & \{ 0.732031\, -0.104875 i \}\\
\hline
     &   0.392421\, -0.152449 i & 0.454183\, -0.152213 i & 0.515138\, -0.152246 i & 0.576192\, -0.152236 i & 0.637204\, -0.152227 i  \\
2    &  (0.410945\, -0.159274 i) & (0.474763\, -0.159321 i) & (0.538728\, -0.159287 i) & (0.602708\, -0.159242 i) & (0.666495\, -0.159241 i) \\
     &  [0.428659\, -0.166365 i] & [0.495658\, -0.166269 i] & [0.562294\, -0.166278 i] & [0.629075\, -0.16623 i] & [0.695693\, -0.1662221 i]\\
     & \{0.45069\, -0.174966 i \} & \{0.52103\, -0.174901 i \} & \{ 0.591224\, -0.174868 i \} & \{ 0.66146\, -0.174815 i \} & \{ 0.731548\, -0.174796 i \}\\
\hline   
\end{tabular}
\label{table:First_set}
}
\end{table*}



\begin{table*} 
\centering
\caption{QN frequencies for electromagnetic perturbations, setting $M=1$ and $\Lambda_0=0.1$ for three different values of the scale-dependent parameter $\epsilon$. For comparison reasons, we also show the frequencies of the classical geometry, i.e., when $\epsilon =0$.
The modes are presented as follows: 
i) classical modes (without parenthesis), 
ii) scale-dependent modes with $\epsilon = 0.01$ (in parenthesis),
iii) scale-dependent modes with $\epsilon=0.015$ (in brackets), and finally
iv) scale-dependent modes with $\epsilon=0.02$ (in curly brackets). 
}
\resizebox{2\columnwidth}{!}{%
\begin{tabular}{cccccc}
\hline
$n$ &  $l=6$ & $l=7$ & $l=8$ & $l=9$ & $l=10$\\
\hline
     &  0.393183\, -0.030421 i & 0.454364\, -0.0304228 i & 0.515465\, -0.0304243 i &  0.576515\, -0.0304253 i & 0.63753\, -0.0304259 i \\
0    &  (0.411284\, -0.0318204 i) & (0.475275\, -0.031823 i) & 0.539191\, -0.0318245 i) & (0.603054\, -0.0318256 i) & (0.666877\, -0.0318264 i) \\
     & [0.429299\, -0.0332147 i] & [0.496101\, -0.0332172 i] & [0.56282\, -0.033219 i] & [0.629482\, -0.0332201 i] & [0.696103\, -0.033221 i]\\
     & \{0.451433	-0.0349266 i\} & \{0.521681	-0.0349295 i\} & \{0.591841	-0.0349315 i\} & \{0.661941	-0.0349329 i\} & \{0.731999	-0.0349339 i\} \\
\hline
     &  0.392931\, -0.091264 i     &   0.454167\, -0.0912645 i    &   0.51527\, -0.0912737 i     &   0.576333 \, -0.0912776 i & 0.63737, -0.0912787 i \\	
1    &  (0.41105\, -0.0954503 i) & ( 0.475014\, -0.095472 i) & (0.538965\, -0.0954752 i) & (0.602855\, -0.0954776 i) & (0.666688\, -0.095481 i) \\ 
     &  [0.428948\, -0.0996512 i] & [0.495802\, -0.0996562 i] & [0.562559\, -0.0996598 i] & [0.629254\, -0.0996619 i] & [0.695893\, -0.0996648 i] \\
     &  \{0.45103\, -0.104787 i\} & \{0.521339\, -0.104793 i\} & \{0.591542\, -0.104797 i\} & \{0.66167\, -0.104802 i \} & \{0.731757\, -0.104804 i \}\\
\hline
     &   0.392453\, -0.1521 i &  0.453873\, -0.15206 i & 0.514898\, -0.152121 i & 0.575956\, -0.152139 i & 0.637055\, -0.152133 i  \\
2    &  (0.410791\, -0.158967 i) & (0.474483\, -0.159133 i) & (0.538515\, -0.15913) & (0.602473\, -0.159128 i) & (0.666297\, -0.1591451 i) \\
     &  [0.428199\, -0.166127 i] & [0.495176\, -0.166117 i] & [0.562026\, -0.166112 i] & [0.628805\, -0.166106 i] & [0.695466\, -0.166115 i]\\
     & \{0.450181\, -0.174687 i \} & \{0.520641\, -0.174673 i \} & \{0.590941\, -0.174673 i \} & \{0.661112\, -0.174683 i\} & \{0.731273\, -0.174679 i \}\\
\hline   
\end{tabular}
\label{table:Second_set}
}
\end{table*}


\begin{table*}
\centering
\caption{Dirac perturbations with $V_+$: Fundamental mode $n=0$ for $\xi=7,8,9,10,11$.}
\begin{tabular}{c | c c c c}
  & \multicolumn{4}{c}{{\sc QN frequencies for Dirac perturbations}} \\
 $\xi$ & $\epsilon=0$ & $\epsilon=0.001$ & $\epsilon=0.002$ & $\epsilon=0.003$ \\
\hline
\hline
7   & 0.4260-0.0304 i  & 0.4277-0.0303 i & 0.4281-0.0304 i & 0.4307-0.0304 i  \\
8   & 0.4868-0.0304 i  & 0.4870-0.0305 i & 0.4879-0.0305 i & 0.4897-0.0306 i  \\
9   & 0.5477-0.0304 i  & 0.5484-0.0304 i & 0.5489-0.0305 i & 0.5510-0.0306 i  \\
10  & 0.6085-0.0304 i  & 0.6092-0.0304 i & 0.6101-0.0305 i & 0.6119-0.0306 i  \\
11  & 0.6694-0.0304 i  & 0.6699-0.0305 i & 0.6711-0.0305 i & 0.6731-0.0306 i  \\
\end{tabular}
\label{table:Third_set}
\end{table*}

\begin{figure*}[ht]
\centering
\includegraphics[width=0.48\textwidth]{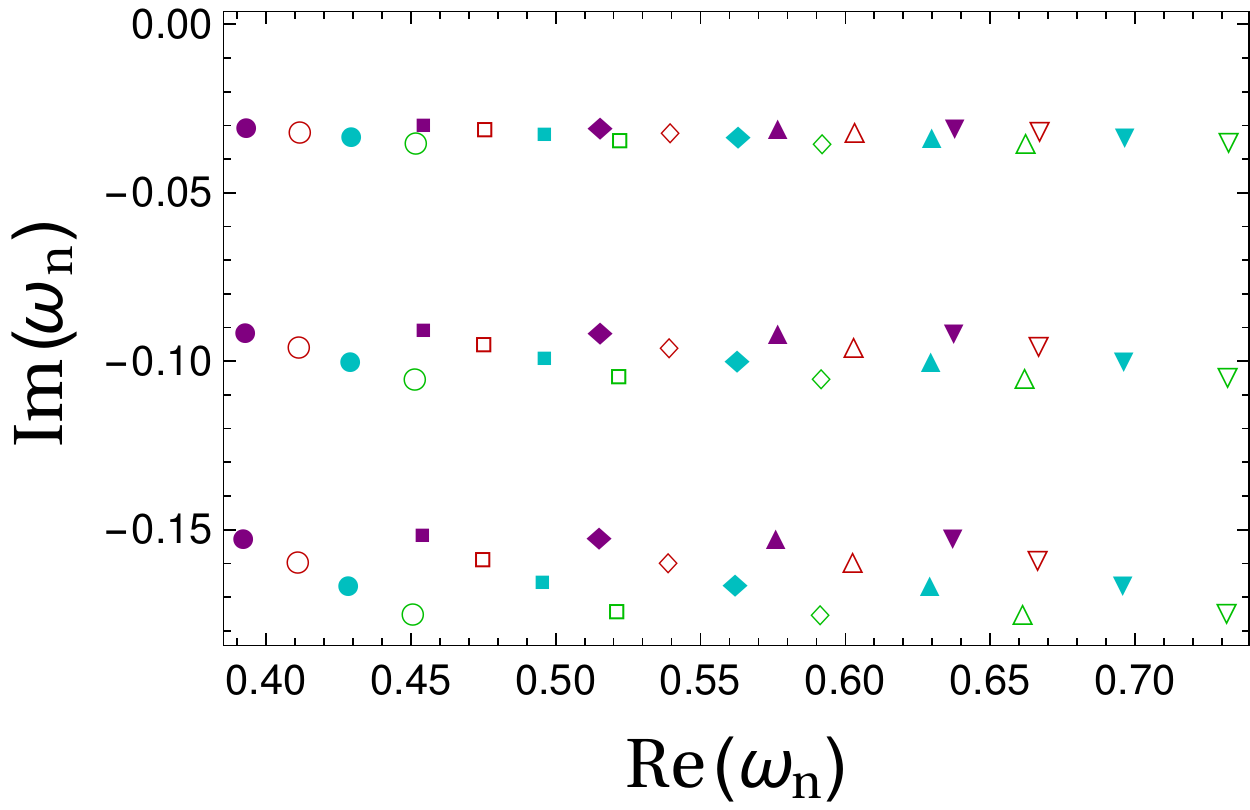}   
\ \ \
\includegraphics[width=0.48\textwidth]{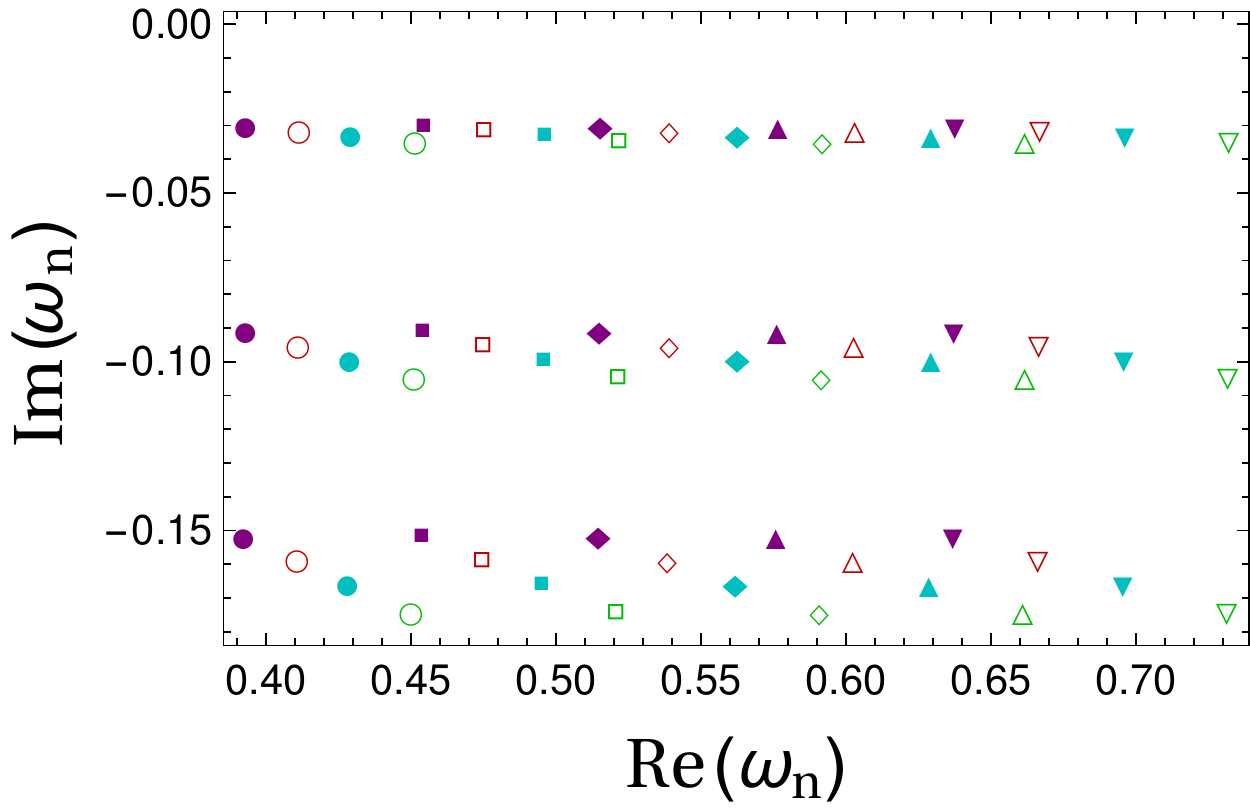}   
\\
\caption{
Quasinormal modes $\omega_n \equiv \omega_R + i \omega_I$. 
{\bf{LEFT:}} (Scalar field) $\text{Im}(\omega_n)$ versus $\text{Re}(\omega_n)$ for $M=1$ and $\Lambda_0=0.1$ in five different cases: 
i)   $l=6$ (points), 
ii)  $l=7$ (squares), 
iii) $l=8$ (rhombus), 
iv)  $l=9$ (triangles), and
 v)  $l=10$ (inverted triangles)
{\bf{RIGHT:}} Same as in the left panel, but for the Maxwell field. Also notice the color code, i.e., 
i) purple for the classical case, 
ii) red for $\epsilon=0.01$,
iii) cyan for $\epsilon=0.015$ and finally,
iv) green for $\epsilon=0.02$. From top to down, we consider from $n=0$ to $n=2$.
}
\label{fig:frequencies_scalar_EM}
\end{figure*}
\begin{figure}[ht]
\centering
\includegraphics[width=0.48\textwidth]{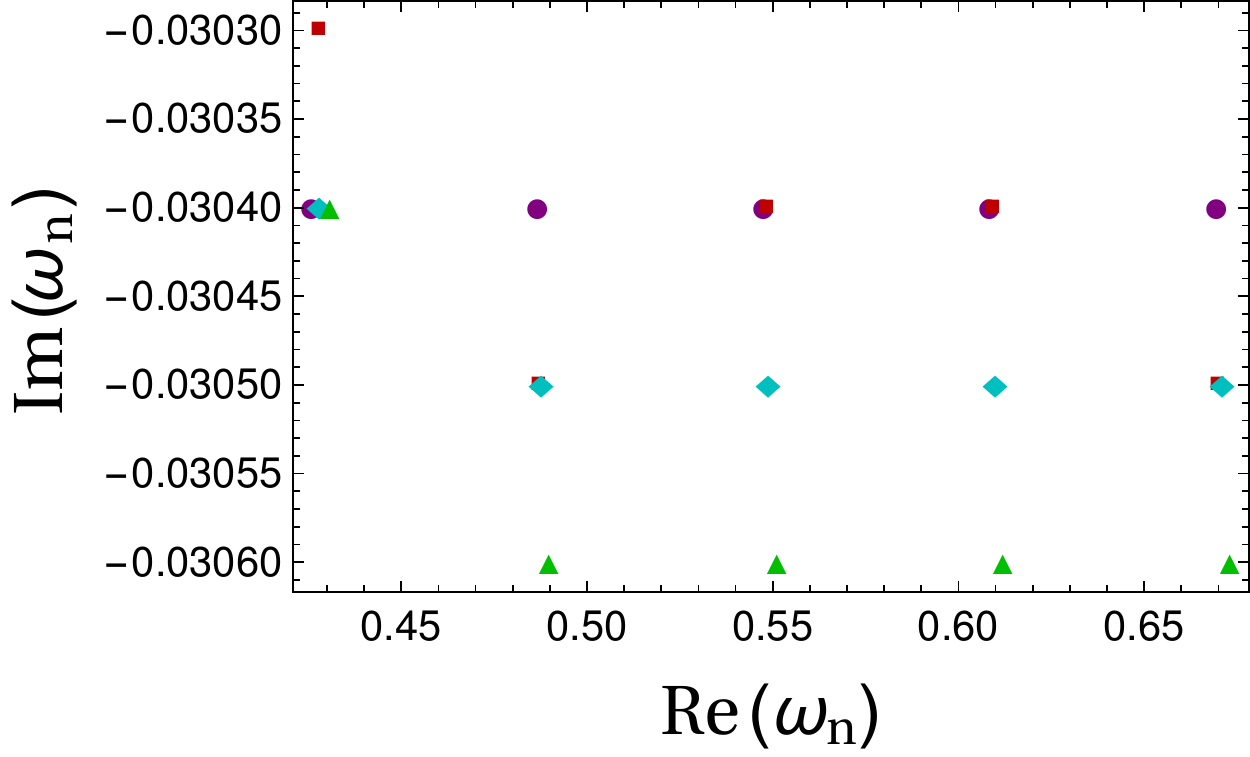}   
\caption{
Quasinormal modes (Dirac field) $\omega_n \equiv \omega_R + i \omega_I$ for the potential with the plus sign, $V_{+}$. $\text{Im}(\omega_n)$ versus $\text{Re}(\omega_n)$ for $M=1$ and $\Lambda_0=0.1$ in five different cases of $\xi$. From left to right, we consider from $\xi=7$ to $\xi=11$.
Also notice the color code, i.e., 
  i) purple points for the classical case, 
 ii) red squares for $\epsilon=0.001$,
iii) cyan rhombus for $\epsilon=0.002$ and finally,
 iv) green triangles for $\epsilon=0.003$. 
}
\label{fig:frequencies_Dirac}
\end{figure}

The obtained QNMs are summarized in Tables \ref{table:First_set}, \ref{table:Second_set} and \ref{table:Third_set}, and they are pictorially shown in Figures~ \ref{fig:frequencies_scalar_EM} and \ref{fig:frequencies_Dirac}. We recall that the information on the stability of the modes is encoded into the sign of the imaginary part of the QN modes. In particular, given the time dependence, $\sim e^{-i \omega t}$, it is easy to verify that when $\omega_I > 0$ the perturbation grows exponentially (unstable mode), whereas when $\omega_I < 0$ the perturbation decays exponentially (stable mode). 
Furthermore, the real part of the modes determines the frequency of the oscillation, 
$\nu = \omega/(2 \pi)$, while the inverse of the absolute value of the imaginary part of the modes determines the damping time, $\tau=1/|\omega_I|$. Those imply that when the real part increases the modes oscillate faster, while when the absolute value of the imaginary part increases the modes decay faster.

\smallskip

Our results show i) that all modes are found to be stable, ii) the SD spectra follow the pattern of the classical spectra regarding the impact of the overtone number and the angular degree, and iii) for given $l,n$ both the real part and the absolute value of the imaginary part of the frequencies increase with the parameter $\epsilon$ that measures the deviation from the classical geometry. Therefore, in the framework of scale-dependent gravity the QN modes are characterized by a lower damping time, they oscillate faster and they decay faster than their classical counterparts. Moreover, modes characterized by a larger damping time will dominate at later times during the quasinormal ringing.

\smallskip

Regarding future work, it would be both challenging and interesting to compute the QNMs for gravitational perturbations, since they may serve as a distinguisher between GR and alternative theories of gravity, see e.g. \cite{Bhattacharyya:2017tyc}. There it was shown that contrary to what happens in GR, where axial and polar modes share equal amounts of radiation, this does not hold in $f(R)$ theories of gravity. Moreover, the strong cosmic censorship conjecture \cite{Penrose:1969pc,Dafermos:2012np} is a hot topic in black hole physics. It would be interesting to see if the conclusions obtained in GR for geometries with a positive cosmological constant (see e.g. \cite{Mellor:1989ac,Dias:2018ynt} and references therein) are respected or violated in SD gravity. We hope to be able to address those issues in a forthcoming article.


\section{Conclusions}

Summarizing our work, in the present article we have studied the quasinormal spectra for massless scalar, Dirac as well as electromagnetic perturbations of four-dimensional scale-dependent Schwarzschild-de Sitter black holes. After reviewing the basic effective field equations of scale-dependent gravity, we perturbed the black hole with a test massless field, and we investigated its propagation into a fixed gravitational background. The wave equations with the corresponding effective potential barrier were presented. The QN frequencies were numerically computed employing the WKB method of 6th order. For better visualization, we showed on the ($\omega_R-\omega_I$) plane the impact on the spectrum of the overtone number, the angular degree and the scale-dependent parameter $\epsilon$. All modes were found to be stable. Our main results indicate that 
i) when $l$ increases, the modes corresponding to the scalar and Maxwell field decreases,
ii) when $\epsilon$ increases, for a given $l$, the modes for the scalar and Maxwell field decrease, and finally
iii) when $n$ increases, for a given $\epsilon$ and $l$, both the real and the imaginary part of all modes decrease.
 Finally, within SD gravity the QN modes oscillate and decay faster in comparison with their classical counterparts.


\section*{Acknowlegements}

We are grateful to the anonymous reviewer for useful comments and suggestions.
The author G.~P. thanks the Fun\-da\c c\~ao para a Ci\^encia e Tecnologia (FCT), 
Portugal, for the financial support to the Center for Astrophysics and 
Gravitation-CENTRA, Instituto Superior T\'ecnico, Universidade de Lisboa, 
through the Project No.~UIDB/00099/2020. 
The author \'A.~R. acknowledges DI-VRIEA for financial support through Proyecto Postdoctorado 2019 VRIEA-PUCV.


\bibliographystyle{unsrt} 
\bibliography{Bibliography_New}  

\end{document}